\documentclass[fleqn,usenatbib]{mnras}

\usepackage{newtxtext,newtxmath}

\usepackage[T1]{fontenc}

\DeclareRobustCommand{\VAN}[3]{#2}
\let\VANthebibliography\thebibliography
\def\thebibliography{\DeclareRobustCommand{\VAN}[3]{##3}\VANthebibliography}

\usepackage{graphicx}	
\usepackage{amsmath}	
\usepackage{enumitem} 

\title[Observing the Underworld]{Observing the Galactic Underworld: Predicting photometry and astrometry from compact remnant microlensing events}

\author[D. Sweeney et al.]{
David Sweeney$^{1, 2, 3}$\thanks{E-mail: david.sweeney@sydney.edu.au}, 
Peter Tuthill$^{1}$, 
Alberto Krone-Martins$^{2}$, 
Antoine Mérand$^{3}$, 
Richard Scalzo$^{4}$, 
\newauthor
\hspace{1mm}Marc-Antoine Martinod$^{1}$
\\
$^{1}$Sydney Institute for Astronomy (SIfA), The University of Sydney, Physics Road, Sydney 2050, Australia\\
$^{2}$Donald Bren School of Information and Computer Sciences, University of California, Irvine, CA 92697, USA\\
$^{3}$European Southern Observatory, Karl-Schwarzschild-Straße 2, 85748 Garching, Germany\\
$^{4}$CSIRO Data61, Clayton, VIC 3168, Australia
}

\date{Accepted XXX. Received YYY; in original form ZZZ}

\pubyear{2023}

\begin{document}
\label{firstpage}
\pagerange{\pageref{firstpage}--\pageref{lastpage}}
\maketitle

\begin{abstract}
Isolated black holes (BHs) and neutron stars (NSs) are largely undetectable across the electromagnetic spectrum. For this reason, our only real prospect of observing these isolated compact remnants is via microlensing; a feat recently performed for the first time. However, characterisation of the microlensing events caused by BHs and NSs is still in its infancy. In this work, we perform N-body simulations to explore the frequency and physical characteristics of microlensing events across the entire sky. Our simulations find that every year we can expect $88_{-6}^{+6}$ BH, $6.8_{-1.6}^{+1.7}$ NS and $20^{+30}_{-20}$ stellar microlensing events which cause an astrometric shift larger than 2~mas. Similarly, we can expect $21_{-3}^{+3}$ BH, $18_{-3}^{+3}$ NS and $7500_{-500}^{+500}$ stellar microlensing events which cause a bump magnitude larger than 1~mag. Leveraging a more comprehensive dynamical model than prior work, we predict the fraction of microlensing events caused by BHs as a function of Einstein time to be smaller than previously thought. Comparison of our microlensing simulations to events in Gaia finds good agreement. Finally, we predict that in the combination of Gaia and GaiaNIR data there will be $14700_{-900}^{+600}$ BH and $1600_{-200}^{+300}$ NS events creating a centroid shift larger than 1~mas and $330_{-120}^{+100}$ BH and $310_{-100}^{+110}$ NS events causing bump magnitudes $> 1$. Of these, $<10$ BH and $5_{-5}^{+10}$ NS events should be detectable using current analysis techniques. These results inform future astrometric mission design, such as GaiaNIR, as they indicate that, compared to stellar events, there are fewer observable BH events than previously thought. 
\end{abstract}




\section{Introduction}

Gravitational microlensing occurs when there is a close alignment between a background source, a foreground object and an observer. The mass of the foreground object warps nearby spacetime and thus the light of the background source is ``bent'' as it travels through this warped region. The end result is that when we observe the background source, the signal has been altered by the presence of the foreground object, i.e. the background source has been \emph{lensed} by the foreground object. This can manifest in several ways: the apparent brightening of the background source, a shift in the apparent position of the background source or in multiple observable images of the background source.

As two celestial bodies participate in microlensing, we can use events to gain information on either or both. 
Since the strength of the lensing depends on the closeness of the alignment between the lens and the background source and the mass of the lens, microlensing events can be used to ascertain not only the presence of a lens, but also its mass. 
This has been done with white dwarfs \citep{Sahu2017} and was recently done for an isolated BH of about 8~M$_\odot$ \citep{Lam2022, Sahu2022, Mroz2022, Lam2023}. Microlensing events caused by isolated BHs and NSs are of particular interest to modern astronomy because these objects are virtually invisible to traditional methods of observation due to their lack of detectable electromagnetic signatures.

The population of compact remnants, BHs and NSs, has not been observationally characterised; a problem that is particularly acute for isolated remnants for which few are confirmed in the literature. Isolated pulsars are readily observable in the radio for an interval after their birth, but with spin-down they fade from view on a (cosmically) brief timescale. Once these pulsars darken, NSs become nearly as invisible as BHs. With the singular exception  mentioned above, all confirmed stellar mass BHs have been identified through their interactions in binaries: as a component in massive X-Ray binaries, radial velocity measurements \citep{El-Badry2023, Nagarajan2023} or through inspiral signals recorded in gravitational wave detectors \citep{GravWave2016}. Outside of binary interactions, which obviously do not occur for isolated remnants, the only current technology capable of detecting these remnants is the use of microlensing measurements. There have been a number of large-scale surveys studying microlensing events: MACHO \citep{MACHO1993}, EROS \citep{EROS1993}, OGLE \citep{OGLE1992} and KMTNet \citep{KMTNet2016}, but while these surveys are sensitive to NS and BH microlensing events, they cannot distinguish them from stellar events because light-curve-only observations cannot break model ambiguity in parameters describing the lens mass, distance and relative proper motion of the event.

The follow-up required for full characterisation of these events is demanding --- the only successfully confirmed BH detection required the use of the Hubble Space Telescope (HST) over 6 years \citep{Lam2022, Sahu2022}. Even with HST data, several reanalyses were required before agreement was reached on the mass of the lensing BH \citep{Mroz2022, Lam2023}. Utilising interferometric follow up is a promising avenue for obtaining astrometric information on these events without the need for years of HST observations, however this method is yet to post a successful detection. Either way, it is crucial that we understand the characteristics of these events so that we may best target them with a coherent, integrated and multi-facility observing strategy \citep[see][]{Gould2023}.

To do this we must model BH and NS microlensing events and compare them to stellar events. 
Previous work has used various stellar population models, which improve with every iteration, to construct a population of compact remnants and calculate microlensing events \citep{Han2003, Wood2005, Oslowski2008, Dai2015, PopSyCLE}. The latest of these, \citet{PopSyCLE} and subsequent \citet{Rose2022}, advanced the field by simulating realistic extinction for high mass stars. 

In this work we take advantage of an improved population synthesis model for compact remnants and use the Gaia catalogue as our source for background stars so that our selection function is more realistic than previous modelling. Our improved lens population synthesis model comes from the work by \citet{Sweeney2022} which, in contrast to PopSyCLE \citep{PopSyCLE} and the work by \citet{Olejak2020}, evolves BHs and NSs post-supernovae. This evolution substantially alters the BH and NS distribution in the Galaxy due to the large kicks ($\sim 450$~km/s for NSs, $\sim80$~km/s for BHs) received by NSs and BHs during their birth in supernovae. We use these populations to calculate microlensing events across the entire sky and compare our results to the previous characterisation by \citet{PopSyCLE} and the microlensing events observed in Gaia \citep{Wyrzykowski2023}. We also make predictions about microlensing events detectable in the combination of Gaia and GaiaNIR \citep{GaiaNIR} data.

The first confirmed detection of a stellar mass BH via microlensing \citep{Lam2022, Sahu2022, Mroz2022, Lam2023} along with the handful of ambiguous detections of compact objects \citep[][]{Mroz2021} demonstrate that we are on the cusp of uncovering the large scale distribution of compact remnants. In order to efficiently build upon this discovery, it is crucial that observing strategies be informed by the accurate parameterisation of the population of compact remnant lensing events. The impending opening of a new observational window on this previously hidden galactic population motivates re-characterisation of microlensing events involving compact remnants, now leveraging the improved distributions of \citet{Sweeney2022}. In turn, distributions of events may be confronted with new data from Gaia and inform forthcoming global astrometry mission proposals such as GaiaNIR.

\section{Data}
\label{sec:data}

\begin{figure}
	\includegraphics[width=\columnwidth]{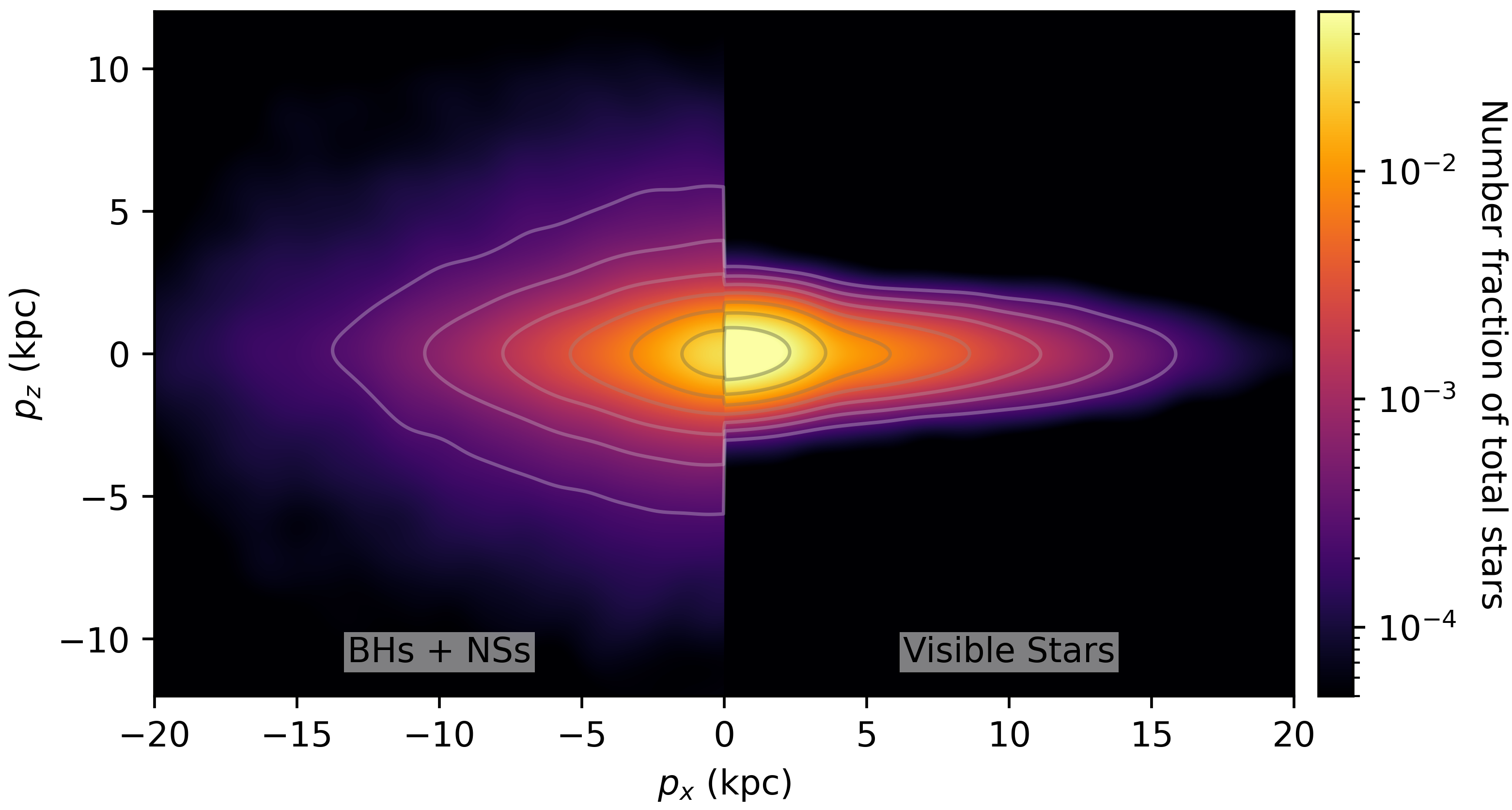}
    \caption{Comparison between the probability density map of the population of NSs + BHs and a synthetic visible Milky Way created by {\tt GALAXIA}. Because of their greater number, NSs dominate the BH + NS distribution. The probability density is  estimated by a gaussian kernel density estimator such that the integral over the whole $x$ -- $z$ plane is 1 and logarithmically spaced contour lines are placed on top of the plot. This plot has been recreated from \citet{Sweeney2022}.}
    \label{fig:GUW}
\end{figure}

There are three sets of data used in this work. The first is the data used to describe the distribution of BGSs which is taken from the Gaia catalogue \citep{Gaia, GaiaDR3}. The sources which populate the catalogue have already been affected by extinction due to dust, observational biases toward luminous objects and crowded field effects. By utilising the Gaia catalogue as BGSs our simulations inherit this selection function, ensuring that the microlensing events simulated are from plausibly monitored sources. Null values in the Gaia catalogue are discussed in Appendix~\ref{app:null-Gaia}. Using the Gaia catalogue has the added advantage that it allows comparisons to be easily made between the predictions given in this work and the observed rate of microlensing in Gaia. 

The second set of data is the set of BH and NS lenses which comes from the paper by \citet{Sweeney2022}. This distribution is reached by starting with a synthetic stellar distribution of the Milky Way generated by {\tt GALAXIA} \citep{Galaxia}.
{\tt GALAXIA} formerly only modelled stellar populations: objects still burning nuclear fuel, but was modified to output stars that evolve past their nuclear burning lifespan. 
Critically, {\tt GALAXIA} synthesises a realistic population which evolves through cosmic time, resulting in a galaxy which has a record of when each remnant formed, allowing the distribution to be modified to account for remnant-specific evolution (e.g. natal kicks).
From this distribution of exhausted stars, NS and BH progenitors were identified based on the initial mass of each progenitor star: 8--$25$~M$_\odot$ were evolved into NSs and $> 25$~M$_\odot$ became BHs \citep{Heger2003, Smartt2009}. Remnants were then given a natal kick drawn from the bimodal Maxwellian distribution published by \citet{Igoshev2020} which was derived from the observed velocity of pulsars. This natal kick was added to the kinematics the remnant inherited from its main sequence progenitor. They key difference from previous work is that each remnant was then evolved through the Galactic potential \citep[{\tt MWPotential2014} from {\tt galpy}; ][]{galpy} for a duration equal to the time since the remnant formed. This process resulted in a remnant population which has a dramatically different spatial distribution compared to the visible Galaxy, as shown in Figure~\ref{fig:GUW}. It should be noted that the data files from \citet{Sweeney2022} undersample the true galactic population by a factor of 1000 and we correct for this during our analysis.

The BH and NS population makes a number of simplifying assumptions, summarised as follows:

\begin{itemize}
    \item The distribution generated by {\tt GALAXIA} is correct. {\tt GALAXIA} was used to generate data based on the Besançon Galaxy \citep{Robin2003} model with some modifications. This model matches observational constraints well but has no history of Galactic mergers. {\tt GALAXIA} also doesn't generate multiple star systems. This is not thought to be a major simplification if considering only isolated BHs and NSs, as we have done here.
    \item Compact remnants are split into BHs and NSs based on the initial mass of their progenitor. Progenitors with initial masses of 8--25M$_{\sun}$ become NSs and progenitors with initial masses >25M$_{\sun}$ become BHs. This simplifying assumption is frequently made \citep[e.g.][]{Fryer1999, Belczynski2008, Fryer2012} and has some observation and theoretical support \citep{Smartt2015, Adams2017}. Metalicity likely also plays a role, but averaged over the entire population we expect this effect to be small.
    \item NSs receive a natal kick following the bimodal Maxwellian derived in \citet{Igoshev2020} from pulsar observations. While the exact formulation of NS natal kicks is not settled, the distribution is thought to be similar to \citet{Igoshev2020} \citep{Lyne1994, Hobbs2005, Katsuda2018}. BHs receive a natal kick with the same momentum, which, due to their larger mass, launches them at lower velocity. BHs with progenitors with initial masses >40M$_{\sun}$ were said to undergo direct collapse and were not kicked. Little is known about BH natal kicks, so this is not the only formulation possible.
    \item BHs all have a mass of 7.8M$_{\sun}$ and NSs all have a mass of 1.35M$_{\sun}$. While there is not significant variance in NS masses \citep{Ozel2012, Postnov2014, Pejcha2015, Sukhbold2016}, there is thought to be a large range of masses for stellar mass BHs. The range and distribution of BH masses is not known and would (in this formulation) impact the natal kicks that BHs receive. 7.8M$_{\sun}$ was chosen because it is a widely accepted median value \citep{Ozel2010, Spera2015, Pejcha2015, Sukhbold2016}.
    \item That the {\tt MWPotential2014} Galactic potential is correct. In practice this is likely not the case as it does not contain a bar which affects microlensing observations toward the bulge \citep{Han1995}, however aside from this deficiency, the {\tt MWPotential2014} potential is a good fit to measured Galactic properties \citep{galpy}. The stellar population does not have to be evolved through the Galactic potential and so still exhibits the bar generated by {\tt GALAXIA}.
\end{itemize}

\noindent
For more details see the original paper \citep{Sweeney2022}.

The third set of data describes the population of stellar lenses which is also generated by {\tt GALAXIA}. The standard release version of {\tt GALAXIA} was used to generate a realistic population of stars in our Galaxy. The stellar data undersamples the true population by a factor of $10^6$, which is also corrected for during our analysis. To convert the absolute magnitudes provided by {\tt GALAXIA} to apparent magnitudes we use the conversion to CTIO V and R filters described in the {\tt GALAXIA} documentation and then use the Johnson-Cousins relationships specified in the Gaia documentation to convert to GaiaDR3 G~magnitudes \citep{GaiaDR3Photometry}:

\begin{align}
    V_\text{CTIO} &= V_\texttt{GALAXIA} + 5\ln\left[\frac{100d}{\text{kpc}}\right] + 3.24 E^{Schlegel}_{B-V}\\
    R_\text{CTIO} &= R_\texttt{GALAXIA} + 5\ln\left[\frac{100d}{\text{kpc}}\right] + 2.634 E^{Schlegel}_{B-V}\\
    G_\text{Gaia} &= V_\text{CTIO} - 0.03088 
                        - 0.04653(V_\text{CTIO} - R_\text{CTIO}) \nonumber\\
                        &\;\;\;\;\;- 0.8794(V_\text{CTIO} - R_\text{CTIO})^2 
                        + 0.1733(V_\text{CTIO} - R_\text{CTIO})^3
\end{align}

Where $d$ is the star's distance from Earth and $E^{Schlegel}_{B-V}$ is the 3D extinction modelled by {\tt GALAXIA} based on the Schlegel maps \citep{Schlegel1998}.

Note that we do not include a population of white dwarfs as they were not modelled in \citet{Sweeney2022}. Modelling microlensing events caused by white dwarfs was done by \citet{PopSyCLE} who found they yield signatures easily distinguishable from BH events, especially when considering microlensing events on the $t_E$--$\pi_E$ plane.

\section{Methods}

\subsection{An introduction to microlensing}
Microlensing, and gravitational lensing more broadly, is a prediction from General Relativity \citep{Einstein1936}. It describes an effect which occurs when there is a close alignment between a massive foreground lens and a background source (BGS). In the case of a point mass, this effect creates two images of the BGS, a bright major image near to the ``true" position of the BGS and a dim minor image on the opposite side of the lens. In the case of an extremely close alignment between the lens and BGS an Einstein ring can appear around the lens instead of a pair of images. While very rare, this Einstein ring gives rise to some natural quantities which describe a lensing event. Firstly there is the Einstein angle which is the angular radius of the Einstein ring \citep{Paczynski1986}. The Einstein angle is typically used as a unit of the scale for microlensing events and is defined as:
\begin{align}
    \theta_E = \sqrt{\frac{4GM_L}{c^2}\frac{D_S - D_L}{D_S\cdot D_L}} 
             = \sqrt{\frac{4GM_L}{c^2 \cdot 1 \text{ pc}}\frac{\bar{\omega}_L - \bar{\omega}_S}{1 \text{ arcsec}}} \label{eq:einstein_angle}
\end{align}
Where $G$ is the gravitational constant, $M_L$ is the mass of the lens, $c$ is the speed of light, $D_L$ and $D_S$ are the distance to the lens and BGS and $\bar{\omega}_L$ and $\bar{\omega}_S$ are the parallax of the lens and BGS, respectively.

The Einstein angle then defines a number of other quantities, such as the separation between the lens and BGS as measure in Einstein angles:
\begin{align}
    \boldsymbol{u} = \frac{\boldsymbol{\theta}}{\theta_E}
\end{align}
Where $\boldsymbol{\theta}$ is the two dimensional vector of the unlensed lens--BGS angular separation. As a note, we follow the convention that bold symbols (e.g. $\boldsymbol{u}$) represent vectors and their unbold variants represent their magnitudes (e.g. $u$). Also note that both $\boldsymbol{\mu}$ and $\boldsymbol{\theta}$ are time dependant since the lens--BGS separation changes with time; sometimes this is made explicit in notation, e.g. by writing $\boldsymbol{u}(t)$. By convention, quantities sub-scripted with 0 (e.g. $u_0$) are the value this quantity takes when the lens and BGS are closest together and the lensing event is strongest.

The timescale of microlensing events is usually related in Einstein times, the time taken for the lens to travel across the Einstein angle:
\begin{align}
    t_E & = \frac{\theta_E}{\mu_\text{rel}}
\end{align}
Where $\mu_\text{rel}$ is the absolute value of the relative proper motion between the lens and BGS.

The creation of two images during a microlensing event leads to two observable quantities of interest: an astrometric shift and a photometric magnification. The astrometric shift of the major and minor images (i.e. the difference in position between the major/minor image and the true position of the BGS) is given by:
\begin{align}
    \delta\boldsymbol{\theta_\pm} = \frac{ \pm\sqrt{u^2 +4} - u}{2} \cdot \frac{\boldsymbol{u}}{u} \cdot \theta_E \label{eq:major-shift}
\end{align}
Where the astrometric shift of the major/minor image is denoted with $\delta\boldsymbol{\theta_+}$/$\delta\boldsymbol{\theta_-}$, respectively. 

In practice this shift is hard to measure as the major and minor images typically are not resolved. Instead we observe the astrometric shift of the centroid due to the combination of the two images:
\begin{align}
    \delta\boldsymbol{\theta_c} = \frac{\boldsymbol{u}}{u^2 + 2} \cdot \theta_E \label{eq:centroid-shift}
\end{align}

This astrometric shift is further complicated by blending between the source and a luminous lens \citep{Dominik2000}. Defining the flux ratio between the lens and source, $g = F_L/F_S$, we can write the blended centroid shift as:
\begin{align}
    \delta\boldsymbol{\theta_{c,LL}} = \frac{\theta_E}{1 + g} \cdot \frac{1+g(u^2 - u \sqrt{u^2 + 4} + 3)}{u^2 + 2 + gu\sqrt{u^2 + 4}} \cdot \boldsymbol{u} \label{eq:blended-shift}
\end{align}
Note that in the case of a non-luminous lens $g = 0$ and this reduces to Equation~\ref{eq:centroid-shift}.

The photometric magnification of the BGS is derived by summing the magnifications of the major and minor image:
\begin{align}
    \mu = \frac{u^2 + 2}{u \sqrt{u^2 + 4}} \label{eq:magnification}
\end{align}

In the case of a luminous lens, the source-lens blending reduces the magnification observed in a photometric light curve, as the luminous lens increases the observed baseline magnitude. To account for this we define a \textit{bump magnitude} which is the difference between the peak magnitude and the baseline magnitude:
\begin{align}
    \Delta m = 2.5 \log_{10} \left(\frac{\mu_0 + g}{1 + g}\right) \label{eq:bump-magnitude}
\end{align}

For a detailed derivation of these quantities see the review by \citet{Mao2008}.

One final quantity is the microlens parallax:
\begin{align}
    \pi_E = \frac{\bar{\omega}_L - \bar{\omega}_S}{\theta_E}
\end{align}
For long microlensing events ($t_E > 100$~days) the microlens parallax can usually be measured from photometry, as the Earth's orbit around the sun imprints an asymmetry into the light curve.

\subsection{Calculating microlensing events}
\label{sec:calculating-lensing}

\begin{figure}
	\includegraphics[width=\columnwidth]{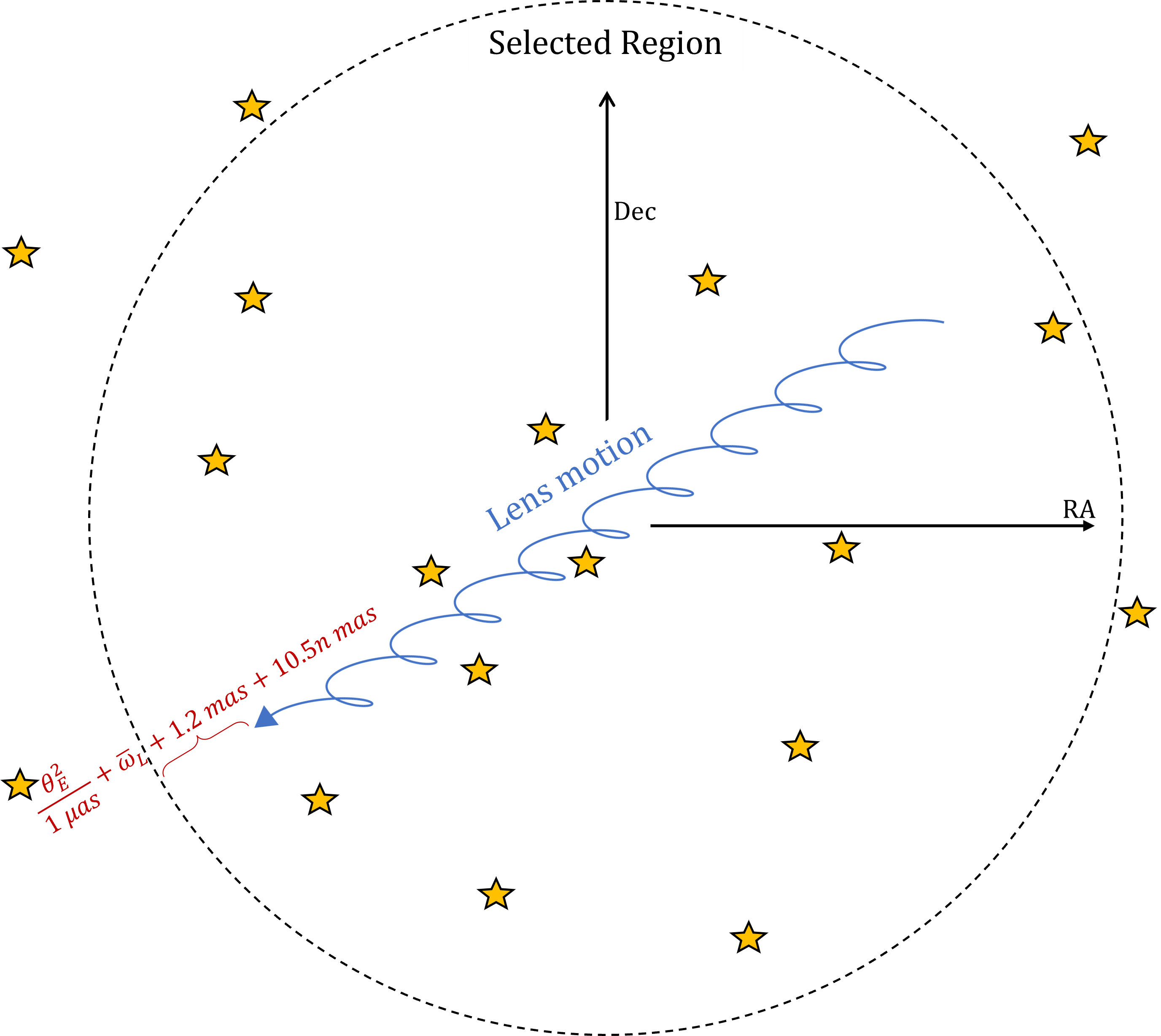}
    \caption{Diagram depicting the motion of a lensing object and the region in which Gaia stars are selected. For each star in this region the microlensing effect caused by the lens will be calculated.}
    \label{fig:selection-cartoon}
\end{figure}

To calculate the number of microlensing events caused by compact remnants (ignoring remnants further than 20~kpc from the Galactic centre) we identified the stars in Gaia which were close enough to each remnant to potentially cause a major image shift larger than 1~$\mu$as. 
Similar to the approach taken by \citet{Kluter2022} for white dwarfs, this was done by calculating the arc of sky traversed by each remnant during the $n$ years of ``observations''. A circle was then drawn which encapsulates this arc with the following buffer term added to the radius:
\begin{align}
    \text{buffer} = \frac{\theta_E^2}{1\,\mu\text{as}} + \bar{\omega}_L + 1.2\,\text{mas} + 10.5n\,\text{mas}
\end{align}
Where the first term on the right-hand side is an approximation of Equation~\ref{eq:centroid-shift} for the maximum angular separation between lens and BGS to result in a 1~$\mu$as astrometric shift. $\theta_E$ is calculated with a BGS parallax of $-1.2$~mas, chosen because 95\% of queried Gaia sources had a larger parallax than this value. The second and third terms are added to encapsulate the parallax motion of the lens and most BGSs. The final term is added to include high proper motion BGSs which might move near to the path. The value of this term is 10.5$n$~mas as 95\% of queried Gaia sources had a proper motion smaller than 10.5~mas/yr. An illustration of the resulting circle is shown in Figure~\ref{fig:selection-cartoon}. It should be noted that because of the chosen parallax and proper motion thresholds our results will slightly underestimate the number of events which occur as we will miss consideration of some BGSs. Testing revealed this underestimation is <1\% for events with a lensing magnification of >1.0001 (i.e. $u \lesssim 20$).

The Gaia catalogue was then queried for stars inside this circle. The returned sources were then filtered to remove objects with a parallax value which was negative by more than 5 times the parallax uncertainty (i.e. $\bar{\omega}_S < 0$ and $|\text{parallax over error}| > 5$) and those with a parallax larger than the lens. Negative parallaxes are unphysical, but removing them entirely has been shown to bias the data \citep{Luri2018}, so we use this filter as a middle ground.

The position, neglecting parallax effects, of the remnant and each star was then calculated over the span of $n$ years. The minimum separation between the two objects was calculated via the haversine formula to within a day. 
A cut on the size of the microlensing event is then applied:
\begin{align}
    \left|\frac{\theta_E^2}{(\theta - (\bar{\omega}_L - \bar{\omega}_S))}\right| \geq 0.8\,\mu\text{as},\\
    \theta - (\bar{\omega}_L - \bar{\omega}_S) < 0
\end{align}
If neither condition is true then the BGS was discarded, otherwise the minimum separation was recalculated with parallax taken into account. The calculation is broken up like this as the minimum separation between two objects without parallax effects is a convex function, and so requires much less computational power to minimise.

The parallax for each object was increased by 1\% to emulate the effect for the Gaia telescope at L2. Taking into account these parallax effects, the minimum separation is then calculated by calculating the minimum separation every day for 100~years either side of the previously found closest approach. If the new minimum separation value is found within 1\% of either extreme of this checked period the period is doubled (respecting the observation bounds) and this process repeats. Finally, the minimum separation is recomputed, checking every 14.4 minutes (0.01 days) within a day either side of the closest approach. This process results in the minimum separation being calculated to within $\sim 15$ minutes.

The minimum separation is then used to calculate the astrometric shifts, photometric magnification and bump magnitude using Equations~\ref{eq:major-shift}--\ref{eq:bump-magnitude}. If the astrometric shift of the major image is smaller than 1~$\mu$as the event is discarded, otherwise details of the remnant, BGS, minimum separation, Einstein angle, Einstein time, major image shift, centroid shift, blended centroid shift, lensing magnification, bump magnitude and the time at which the event occurred are recorded. Since the centroid shift (Equation~\ref{eq:centroid-shift}) is maximised at $u = \pm \sqrt{2}$, if the minimum separation has $u_0 < \sqrt{2}$ the centroid shift is calculated with $u = \sqrt{2}$ as the lens will cause this shift at some point in time. The $u$ which maximises the blended centroid shift depends on $g$. This $u$ is calculated for each event and if $u_0$ is smaller than this value, the maximum blended centroid shift is used. In the case of a remnant that does not cause any major image shifts larger than 1~$\mu$as, its details are recorded with null values in the remaining columns. The 1~$\mu$as major image astrometric cut was chosen to be well below the threshold of observable events.

This results in a list of microlensing events with major image astrometric shifts larger than 1~$\mu$as caused by the compact remnant population derived by \citet{Sweeney2022}. In practice this procedure is parallelised (this is easily done as the effects of each remnant are independent) and the results are combined at the end. 

\subsection{Different timespans}
\label{sec:timespans}

In this work three different simulations, each with a distinct timespan, were performed:

\begin{description}
    \item \emph{Simulation 1} spanned 10~000~years for which BHs and NSs constituted the lensing population. As noted in Section~\ref{sec:data}, the compact remnant population is undersampled by a factor of 1000, so this simulation provides an effective 10 years of compact remnant lensing in the Milky Way.
    \item \emph{Simulation 2} spanned 100~000~years, modelling the scenario where stars form the lensing population. Similar to the BH and NS data, the stellar data is undersampled, now by a factor of $10^6$; therefore this simulation is equivalent to 1/10th of a year of real microlensing observation.
    \item \emph{Simulation 3} modelled a 32 year timespan, once again using BHs and NSs as lenses. This timespan consisted of 12 years of ``observation", where the microlensing events were calculated, a 15-year gap and then 5 years of further observation. This format was chosen to mimic the lifespan of Gaia and then a hiatus of observation before a potential 5-year future GaiaNIR mission. To deliver robust statistics despite the undersampled population, 101 different 32-year timespans were modelled. This was achieved by using the publicly available code from \citet{Sweeney2022} to evolve the compact remnant distribution from its nominal initial (``present day'') configuration by 20~000 years into the future, saving the distribution every 200 years. Each timestep forms an effectively independent lens population for the purposes of the process described in Section~\ref{sec:calculating-lensing} as 99.9\% of lensing events have Einstein times less than 37 years. 
\end{description}

The first and final year of any observation period exhibits an increased number of microlensing event ``detections'' due to events which peak outside of the observing period. This is an important effect for \textit{Simulation 3} as these events have been detected in Gaia data \citep{Wyrzykowski2023}, but is undesirable for \textit{Simulations 1} and \textit{2} which are used to calculate yearly event rates. To remove this issue the first and last year was removed for \textit{Simulations 1} and \textit{2}; the slightly reduced observational periods taken into account in all further analysis. For convenience and brevity we continue to refer to these simulation timespans as 10~000 or 100~000 years of data.

Uncertainties were calculated by computing a 95\% confidence interval via bootstrapping. To perform this bootstrapping, \textit{Simulations 1} and \textit{2} were broken into yearly observations and the various event rates (see Section~\ref{sec:results}) were computed. These quantities were then sampled and combined to provide $10^5$ bootstrap samples. The 2.5th and 97.5th percentiles was then calculated for each statistic, becoming the confidence interval. In the case of \textit{Simulation 3}, it already exists as 101 32-year timespans so these were sampled from instead of the yearly divisions made for \textit{Simulations 1} and \textit{2}.

\section{Results}
\label{sec:results}

\subsection{Expected microlensing events in a year}
\label{sec:yearly}
\begin{figure}
	\includegraphics[width=\columnwidth]{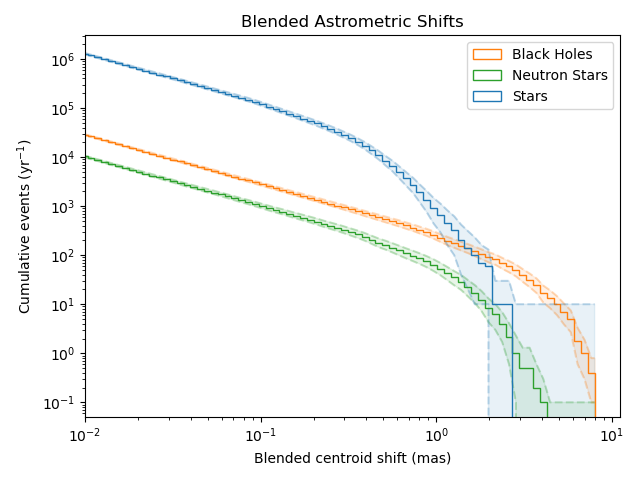}
	\includegraphics[width=\columnwidth]{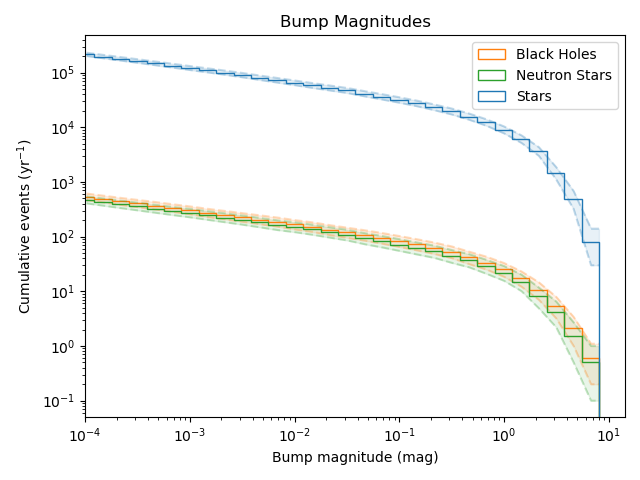}
	\includegraphics[width=\columnwidth]{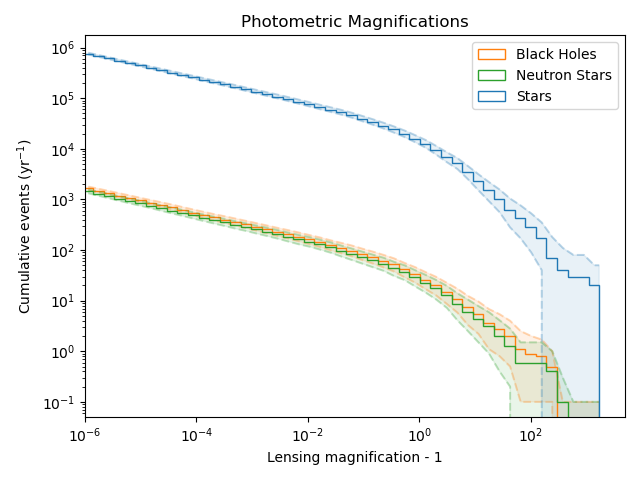}
    \caption{Expected yearly distribution of blended astrometric shifts, bump magnitudes and photometric magnifications caused by microlensing events where the lenses are stars, BHs and NSs. The shaded regions indicate $3\sigma$ uncertainty or the undersampling factor if the histogram bin is empty. Note that in the magnification plot the $x$ axis is the magnification minus one (enabling a log scale can be used). Also note that the granularity of the simulation affects the plots at the right-hand side --- the stellar data only spans 1/10th of a year and so (after correction with the multiplicative factor of 10) the line falls as it reaches $\sim$10 events per year. A more fine-grained simulation would show the stellar relationship smoothly continuing below the 10$^{1}$ threshold. Similarly, the BH and NS lines fall at $\sim$0.1 events per year as the data accounts for 10 years of microlensing events.}
    \label{fig:shifts-and-mags}
\end{figure}

The expected annual distributions of astrometric shifts, bump magnitudes and photometric magnifications caused by microlensing events is given in Figure~\ref{fig:shifts-and-mags}. These plots were generated by taking the data set reached by \textit{Simulation 1} and \textit{Simulation 2} and transforming them into a yearly distribution based on the undersampling factor of the original distribution --- 10$^3$ for NSs and BHs, 10$^6$ for stars. These results indicate that every year we can expect $88_{-6}^{+6}$ BH, $6.8_{-1.6}^{+1.7}$ NS and $20^{+30}_{-20}$ stellar microlensing events which cause a blended astrometric shift larger than 2~mas ($231_{-9}^{+10}$,  $55_{-5}^{+5}$ and $690_{-160}^{+170}$ larger than 1~mas, respectively). Similarly, we expect $21_{-3}^{+3}$ BH, $18_{-3}^{+3}$ NS and $7500_{-500}^{+500}$ stellar events causing bump magnitudes $> 1$~mag as well as $5.5_{-1.4}^{+1.5}$, $4.3_{-1.2}^{+1.3}$ and $1600_{-200}^{+300}$ events with bump magnitudes larger than 2.5. In terms of unblended photometric magnification $26_{-3}^{+3}$ BH, $23_{-3}^{+3}$ NS and $12800_{-700}^{+700}$ stellar events cause magnifications of $>2$ (i.e. a doubling in brightness) as well as $5.5_{-1.4}^{+1.5}$, $4.3_{-1.2}^{+1.3}$ and $2300_{-300}^{+300}$ events with magnifications larger than 10.

\begin{figure}
	\includegraphics[width=\columnwidth]{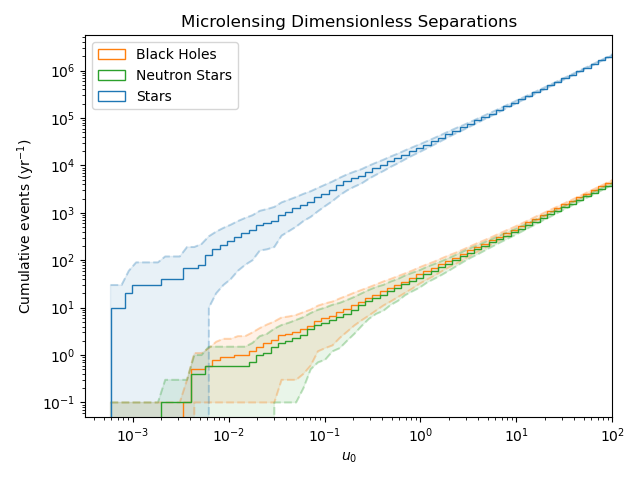}
	\includegraphics[width=\columnwidth]{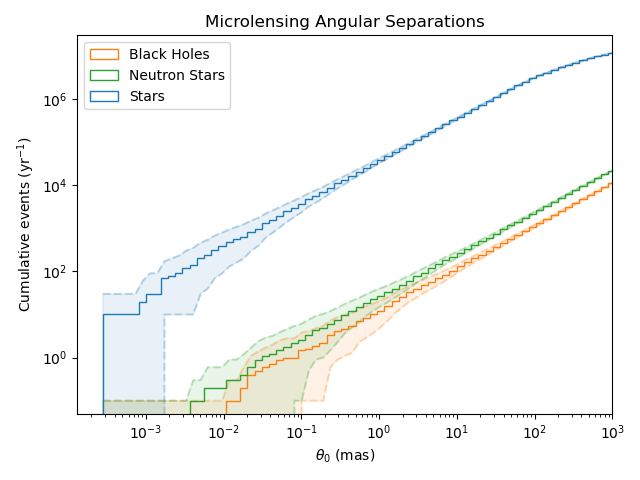}
    \caption{Expected yearly distribution of dimensionless separations (top panel) and angular separations (bottom panel) caused by microlensing events where the lenses are stars, BHs and NSs. The shaded regions indicate $3\sigma$ uncertainty or the undersampling factor if the histogram bin is empty. Note that the stellar data only accounts for 1/10th of a year and so is multiplied by 10, which is why the line falls as it reaches $\sim$10 events per year. The stellar line likely continues below the 10$^{1}$ threshold and is just not picked up by our simulation. This figure includes events down to a major image shift of $1\mu$as, many of which will not be observable.}
    \label{fig:separations}
\end{figure}


These results can be generalised to other all-sky surveys with similar selection functions to Gaia, by scaling the number of stars surveyed in comparison to Gaia, which encompasses approximately 2 billion stars.

\subsection{Characterisation of microlensing events}
\begin{figure}
	\includegraphics[width=\columnwidth]{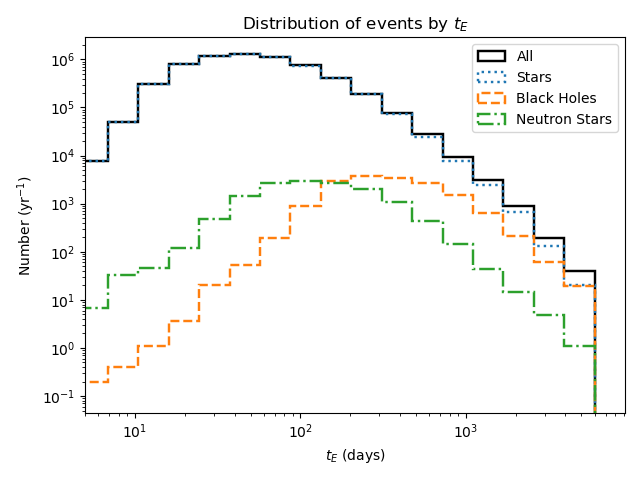}
	\includegraphics[width=\columnwidth]{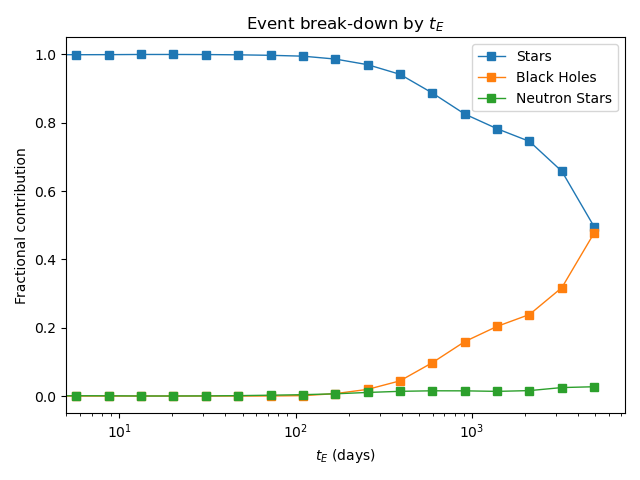}
    \caption{Expected yearly distribution of microlensing events as a function of Einstein time and type of lens. The largest bin/point in this figure was set to be $6\,000$~days so the larger stellar undersampling doesn't mislead the viewer. BH, NS and some stellar events continue beyond this point. This figure includes events down to a bump magnitude of $10^{-10}$~mag, many of which will not be observable.}
    \label{fig:einstein-times}
\end{figure}

The characterisation of microlensing events, particularly by Einstein time, is important for identifying which events to follow-up to efficiently probe the population of BHs and NSs. In particular, identifying events likely to be caused by BHs is of great interest and furthermore their more extreme properties renders them easier to detect. Figure~\ref{fig:einstein-times} shows the distribution of microlensing events as a function of Einstein time split by the type of lens. 

These figures may be directly contrasted with Figure~7 from \citet{PopSyCLE} which, using a different methodology, modelled microlensing events toward the Galactic Bulge. 
Our results show that BHs make up a significantly smaller fraction of long Einstein time events than the 50\% of $t_E > 120$~days that was predicted in \citet{PopSyCLE}.
The discrepancy between the two studies likely arises due to differences in the dynamical model: in the application of the natal kick and subsequent evolution of the remnant. In addition, the two methodologies take divergent approaches to arrive at their populations of BGSs: an augmented {\tt GALAXIA} distribution for the case of \citet{PopSyCLE} and Gaia sources in the case of this work.

In modelling lensing BHs, the \citet{PopSyCLE} work applies natal kicks of 100~km/s with no subsequent evolution through the Galactic gravitational potential over cosmic time. By contrast, the \citet{Sweeney2022} work applies natal kicks which follow a bimodal Maxwellian distribution which, in the case of BHs, has expected value of $\sim77$~km/s. The BHs are then evolved through the Galactic gravitational potential for the timespan from birth to the present day. This acts to reduce the average speed of the BHs while expanding their distribution to larger radii. The end result is that the BHs from \citet{PopSyCLE} have a much larger average speed and are more concentrated towards the galactic centre than \citet{Sweeney2022}: features which decrease the Einstein time and increases the event rates due to BHs. 

This comparatively lower fraction of BHs for long Einstein time events renders the strategy of identifying likely BH events using photometric lightcurves alone much more difficult.  
Analysis of photometric surveys which leveraged the \citet{PopSyCLE} finding that 50\% of events with $t_E \geq 120$~days are BHs may therefore significantly overestimate the number of events ascribed to BH lensing (e.g. \citet{Golovich2022}'s reanalysis of OGLE-III and -IV events). 
The detection of an isolated BH via microlensing \citep{Lam2022, Sahu2022} with an Einstein time of $\sim 250$~days constitutes a reasonably unlikely outcome under our modelling assuming a small number of attempts. However, without knowing the number of events monitored or the selection criteria used it is impossible to draw any conclusions on how this detection should impact the modelling. The characteristics of the event itself were also quite unusual: the magnification was $\sim 400$ which is much larger than the typical event, and the lens distance was found to be $\sim1.6$~kpc, which is unusually close (as shown in Figure~\ref{fig:lens-distance}).

In the case of NSs, \citet{PopSyCLE} apply a natal kick of 350~km/s as opposed to the bimodal Maxwellian distribution with expectation of $\sim450$~km/s in \citet{Sweeney2022}. The evolution through the Galactic potential in \citet{Sweeney2022} leads to a major reduction in speed and an increase in the scale height of the NSs, as shown in Figure~\ref{fig:GUW}. In light of these differences, it's quite surprising that the distributions of microlensing events caused by NSs are relatively similar between the two studies. The similarity may be partially explained by the opposing nature of the two effects: the reduced speed increases the Einstein time of events, but the inflated distance to NSs decreases it. 

The NSs' migration to larger scale heights and slower speeds should dramatically reduce the number of events compared to that of \citet{PopSyCLE}. This study finds $\sim3$ times fewer NS events in ratio to BH events compared to $\sim$an order of magnitude fewer from the \citet{PopSyCLE} work. This cannot be explained by differing numbers of BHs and NSs in the two underlying populations. The \citet{PopSyCLE} population contains $2\times$ more NSs than BHs. In comparison, the \citet{Sweeney2022} start with $4.8\times$ more NSs, but the evolution of the remnants through the Galactic potential \citep[which is not done in][]{PopSyCLE} acts to kick 40\% of NSs out of the Galaxy entirely. The remaining population then contains $2.3\times$ more NSs than BHs, if remnants $> 20$~kpc of the Galactic centre are excluded. This agreement in population make-up combined with the increased speed of the \citet{PopSyCLE} NS population should result in many more NS events being recorded in their counts. However, their low rate of NS events is likely due to a separate model factor entirely: their simulations lack sensitivity to short events which are much more likely due to NSs than BHs. This can be traced to the \citet{PopSyCLE} simulated ``observational cadence'' which is only every 10 days: a span that can easily miss many of these short events, and most particularly those additional events generated by the large velocities of their NS population.

\begin{figure}
	\includegraphics[width=\columnwidth]{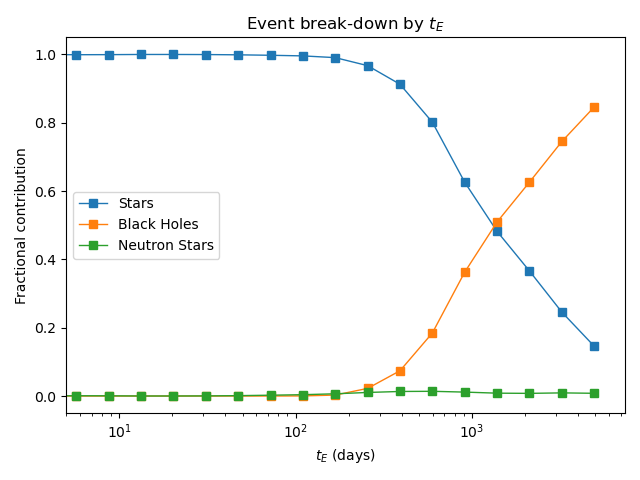}
    \caption{Expected fractional distribution of microlensing events in a scenario where BHs have a mass of $20$~M$_\odot$. The largest bin/point in this figure was set to be $6\,000$~days so the larger stellar undersampling doesn't mislead the viewer. BH, NS and some stellar events continue beyond this point. This figure includes events down to a bump magnitude of $10^{-10}$~mag, many of which will not be observable.}
    \label{fig:20M-times}
\end{figure}

Unsurprisingly, the fractional contributions seen in Figure~\ref{fig:einstein-times} depend on the masses of the lenses involved. For example, if BHs have a mass of $20$~M$_\odot$ (instead of $7.8$~M$_\odot$) but still receive the same natal kick velocity then the fractional contribution as a function of Einstein time is shown in Figure~\ref{fig:20M-times}. This plot shows that the increased BH mass causes microlensing events to become dominated by BHs at long Einstein times, but even with this extreme mass BHs are only a few percent of events with $t_E \approx 250$~days. Event rates are given for this example in Appendix~\ref{app:20M-BH}.

\begin{figure}
	\includegraphics[width=\columnwidth]{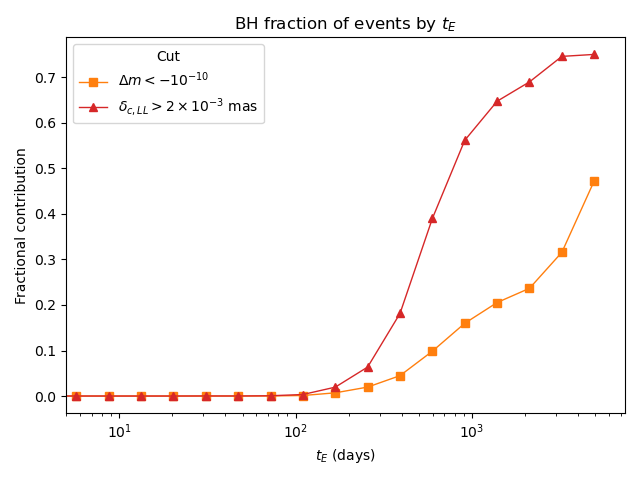}
    \caption{Fraction of events caused by BHs as a function of Einstein time for a bump magnitude and an astrometric cut. The orange line has the same cut applied as the orange line in Figure~\ref{fig:einstein-times} while the red line instead applies an astrometric cut.}
    \label{fig:time-cuts}
\end{figure}

The fractional contributions strongly depend on the selection criteria for events. Figure~\ref{fig:time-cuts} shows how the BH contribution varies for a photometric or astrometric cut. The threshold of this cut does not significantly alter the shape, as shown in Appendix~\ref{app:fraction}. Whether the events are located inside or outside the bulge only makes a small difference to the BH fraction, as shown in Appendix~\ref{app:bulge}. For microlensing events identified via photometric surveys, the flatter curve of BH contribution is expected. The fraction of BH events is much larger if a selection is made based on astrometric criteria, however surveying the sky for small deviations caused by microlensing events is not currently performed.

\begin{figure}
	\includegraphics[width=\columnwidth]{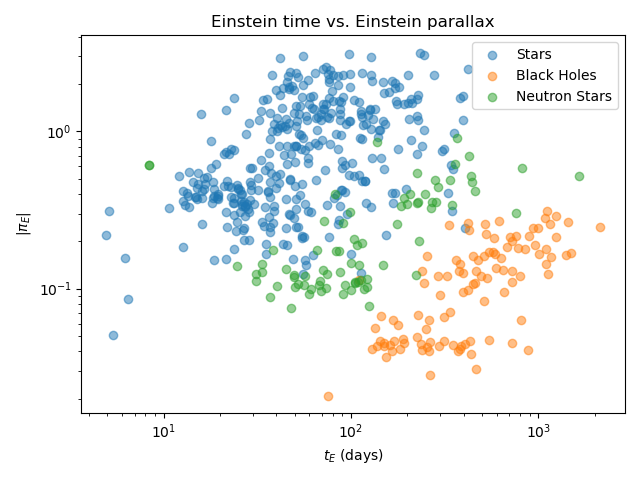}
    \caption{Scatter plot showing the splitting of stellar, BH and NS events across the Einstein time--Einstein parallax plane. Note that there is a factor of 100 difference in the undersampling between the stellar events and the BH and NS events which is not accounted for in this figure. This figure only includes events with a bump magnitude $> 2$~mag.}
    \label{fig:microlens-parallax}
\end{figure}

Instead of solely relying on the Einstein time as a predictor for lens type, it is much better to predict based on both the Einstein time and the Einstein parallax. We plot the scatter of lensing events with a bump magnitude $>2$ in Figure~\ref{fig:microlens-parallax}. If natal kicks are larger than those applied in this work then the BH and NS events would migrate to smaller Einstein times, compressing the $x$-axis of the figure. As discussed by \citet{PopSyCLE} and \citet{Gould2023}, the $t_E$--$\pi_E$ basis splits lensing events by the mass of the lens, with equal mass contours being approximately diagonal. Given the difficulty in identifying BH microlensing events by Einstein time alone, shown in Figure~\ref{fig:einstein-times}, these results suggest it is crucial to also measure the microlens parallax. It should be noted that while BHs and NSs are reasonably distinct in Figure~\ref{fig:microlens-parallax}, they are significantly less separate than in the scatter published by \citet{PopSyCLE}.

\begin{figure}
	\includegraphics[width=\columnwidth]{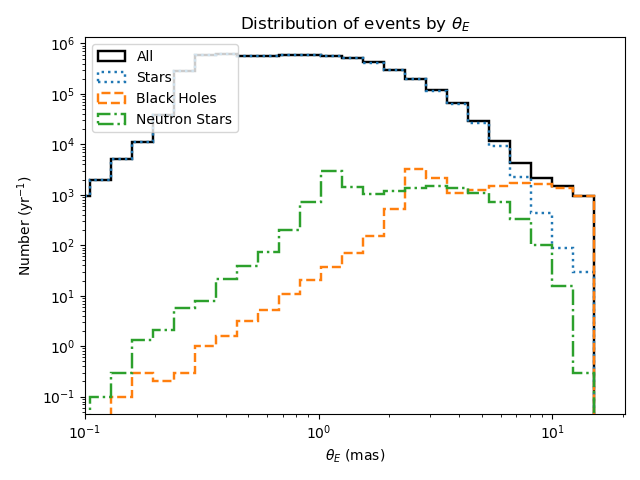}
    \caption{Expected yearly distribution of microlensing events as a function of Einstein angle and type of lens. The largest bin in this figure was set to be $15$~mas so the larger stellar undersampling doesn't mislead the viewer, BH events continue beyond this point. This figure includes events down to a bump magnitude of $10^{-10}$~mag, many of which will not be observable.}
    \label{fig:einstein-angles}
\end{figure}

The distribution of Einstein angles for microlensing events is shown in Figure~\ref{fig:einstein-angles}. We can see from this figure that microlensing events with Einstein angles larger than $\sim$8~mas are mostly caused by BHs.

\begin{figure}
	\includegraphics[width=\columnwidth]{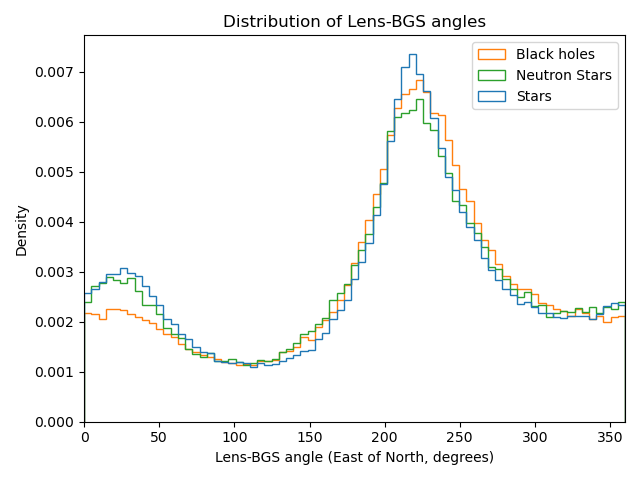}
    \caption{Plot showing the lens-BGS position angle a year in advance of the microlensing event, split by the type of lens.This figure includes events down to a bump magnitude of $10^{-10}$~mag, many of which will not be observable.}
    \label{fig:position-angle}
\end{figure}

A curious result is that the lens-BGS position angle (i.e. the angle, east of north, between the lens and BGS) a year in advance of the microlensing event is not uniformly distributed, as shown in Figure~\ref{fig:position-angle}. We suggest the non-uniformity in this distribution is due to the net proper motion observed in the sky as a result of our motion through the Galaxy. 

In Appendix~\ref{app:lens-info} we provide plots of the lens masses ($M_L$) and lens distances ($D_L$) broken down by type of lens. These are included so that they may be used as priors for observation analysis. 

\subsection{Comparison to Gaia}

\begin{figure}
	\includegraphics[width=\columnwidth]{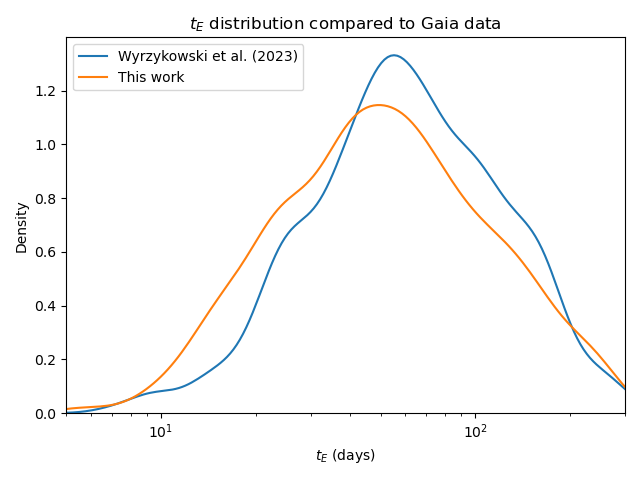}
	\includegraphics[width=\columnwidth]{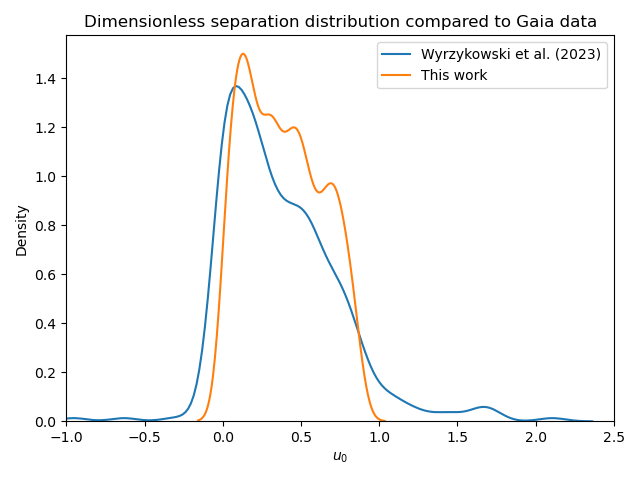}
    \caption{Comparison between the distribution of Einstein times (top panel) and dimensionless separations (bottom panel) found in Gaia data \citep{Wyrzykowski2023} and those reached in this work. 
    The distributions of Einstein times and dimensionless separation show good agreement between this work and the microlensing events found in the Gaia data.
    Microlensing events due to stars dominate the distributions from this work.
    The distribution from this work only includes events with a bump magnitude $>0.4$~mag in order to emulate a photometric selection criteria. The Gaia data plotted is the ``Level 1'' solutions derived in \citet{Wyrzykowski2023}. Some outliers from the Gaia data are beyond the bounds of this plot. The curves were generated using a Gaussian kernel density estimator, so each curve will integrate to 1.}
    \label{fig:Gaia-data}
\end{figure}

We can also compare the predictions of this work to the microlensing events observed in Gaia photometry. \citet{Wyrzykowski2023} analyse the Gaia photometry data from 2014 to 2017 for evidence of microlensing events. They find evidence of 363 microlensing events, of which 163 are identified based on Gaia data alone and the other 200 are confirmed to be present in the Gaia data after a reexamination due to the events being reported in the OGLE-IV survey \citep{Mroz2019, Mroz2020} or the All-Sky Automated Survey for Supernovae \citep[ASAS-SN; ][]{Shappee2014}. This inclusion biases the sample due to the increased detection efficiency towards the bulge where OGLE-IV survey fields lie. Gaia struggles in crowded regions, such as the bulge, and so this positive bias acts to offset the reduced native detection efficiency due to Gaia's sampling. 

To compare the number of events in Gaia data with our simulation we impose cuts to emulate \citet{Wyrzykowski2023}'s selection criteria. We filter both sets of data to only look at the bulge (Galactic longitude $|l| < 10$~degrees, where OGLE assists the most with candidate selection), only consider $< 19$ G magnitude BGSs and events with Einstein times 20~days $ < t_E <$ 250~days. With this criteria, we find that there are $100$ microlensing events per year with bump magnitudes $>3$~mag, of which $<0.1$ are due to BHs and $0.1$ due to NSs. In comparison, \citet{Wyrzykowski2023}'s ``Level 1'' analysis of Gaia data finds there to be $\sim50$ microlensing events per year meeting the above criteria, in reasonable agreement with our simulations. This suggests that it is quite unlikely that many BHs can be found amongst the events identified by \citet{Wyrzykowski2023}. Note that the cuts we apply don't take into account binaries, variable stars or blending with neighbouring stars, all of which would make microlensing detections much more difficult. 

We can also compare the distribution of Einstein times and dimensionless separations between the bright events (bump magnitude $> 0.4$) in this work and the distributions from events in Gaia. This comparison is shown in Figure~\ref{fig:Gaia-data}. It shows that the distribution of Einstein times and dimensionless separations in this work and Gaia data are in good agreement. 
Our bump magnitude threshold was chosen so that the ``plateau'' of the dimensionless separations from this work approximately matches peak width from the Gaia data. Decreasing or increasing this threshold simply shortens or lengthens the plateau as events with greater separations are removed or included. We also plot the distribution of Einstein times for the bulge/non-bulge population in Appendix~\ref{app:bulge}.

\begin{figure}
	\includegraphics[width=\columnwidth]{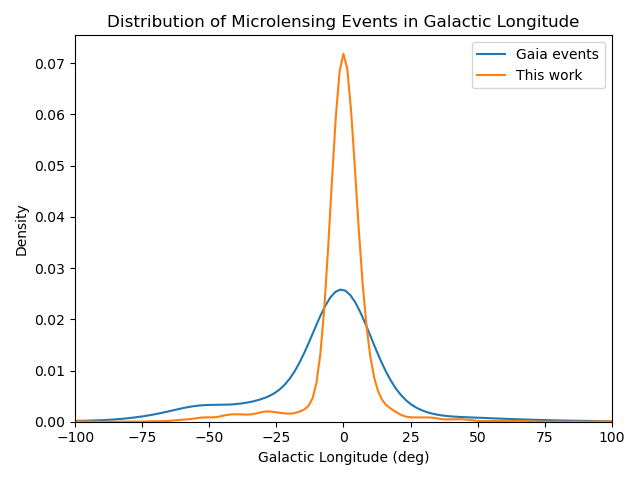}
	\includegraphics[width=\columnwidth]{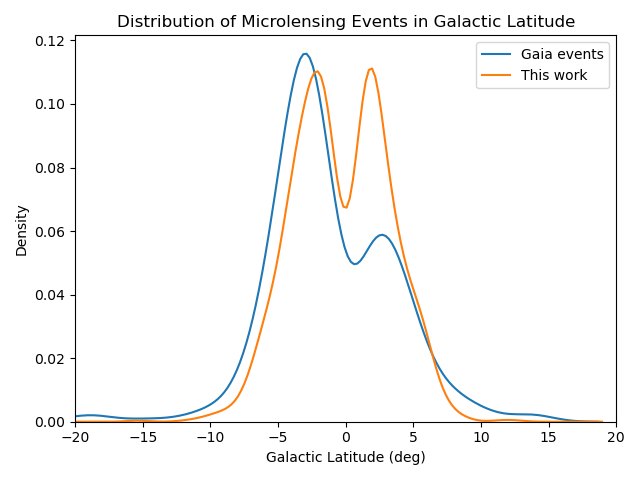}
    \caption{Comparison between the distribution of microlensing events over galactic longitude (top panel) and galactic latitude (bottom panel) found in Gaia data \citep{Wyrzykowski2023} and those reached in this work. 
    Overall, the distribution of event locations shows good agreement between the distributions. 
    Microlensing events due to stars dominate the distributions from this work.
    The distribution from this work only includes events with a bump magnitude $>0.4$~mag in order to emulate a photometric selection criterion. The curves were generated using a Gaussian kernel density estimator, so each curve will integrate to 1.}
    \label{fig:event-locations}
\end{figure}

The locations of the events detected in Gaia data can also be compared to the locations of events from this work. Figure~\ref{fig:event-locations} shows this comparison between distributions. It shows that the locations of events largely agree, but that this work predicts more events toward the bulge than are seen in the Gaia data. This could be due to Gaia's inefficiencies in making measurements in crowded stellar fields, such as the bulge, and suggests losses are not entirely offset by previously mentioned gains in candidate selection from photometric surveys such as OGLE-IV. Leaving aside difficulties in crowded stellar fields, Gaia's detections are also impacted by its scanning law \citep{Gaia}, with the bulge being one of the areas least frequently revisited. The scanning law also explains the positive-negative longitude asymmetry for the Gaia data. During the period considered by \citet{Wyrzykowski2023}, the region between $-15$ and $-50$ longitude was typically scanned $2$--$3\times$ more frequently than its positive counterpart. As discussed in \citet{Wyrzykowski2023}, a reduction in the number of data points and sampling frequency reduces the number of detections that can be made on the basis of Gaia photometry, explaining the larger number of events identified by \citet{Wyrzykowski2023} at negative longitudes.

The lower panel of Figure~\ref{fig:event-locations} shows that our simulations yield two distinct peaks whereas the Gaia data has one major and one minor peak. The peak at a Galactic latitude of $\sim -2$ is a close match between both sets of data. Both data sets then dip at 0 degrees, likely due to Gaia failing to observe stars in the dusty Galactic plane. At $\sim 2$ degrees our data exhibits a peak mirroring that at $\sim -2$, which makes sense --- intuition would have the microlensing event distribution be approximately symmetrical around the Galactic plane. The Gaia data exhibits a much smaller peak in this region. \citet{Wyrzykowski2023}'s analysis is heavily reliant on candidates provided from OGLE-IV to detect events in near the Galactic plane. OGLE-IV samples the $\sim 2$ latitude region much less frequently than the $\sim -2$ region, so \citet{Wyrzykowski2023}'s asymmetric peaks are likely a selection effect arising from their reliance on OGLE.

\subsection{Comparison to OGLE-IV}

If we examine the most monitored OGLE-IV fields (i.e. those monitored $> 10$ time per night: 500, 501, 504, 505, 506, 511, 512, 534 and 675) and approximate the fields to be $64 \times 64$ arcminute squares aligned with the right ascension and declination axes, our results predict $180$ microlensing events per 5 month period. By comparison, the OGLE early warning system \citep{Udalski2015} reported 134 events which peaked during the 5 month 2023 Galactic Bulge season --- April to August --- and had bump magnitudes $> 1$~mag and were located in those fields. This comparison is imperfect: Gaia observes in the visible, while OGLE-IV is conducted in I band and their completenesses toward the bulge vary. In addition, the OGLE-IV survey is conducted at a 1.3m ground based telescope with typical seeing of $1.3$" \citep{Udalski2015} and so is much more susceptible to neighbouring stars contributing to the base magnitude of an event but not participating in the lensing event --- acting to reduce the apparent number of $> 1$~mag events. On the other hand, we don't model white dwarfs, which have been found to constitute $\sim 10$\% of microlensing events towards the bulge \citep{PopSyCLE}. Despite these caveats, this comparison shows that our numbers are at least broadly compatible with OGLE-IV observations. 

\subsection{Predictions for future Gaia data releases}

Photometric microlensing events should continue to be found in Gaia data at the same rate as they were detected in \citet{Wyrzykowski2023}. This rate changes throughout the study, but using the penultimate year of observation (to avoid edge effects) we predict 126 microlensing events per year. Thus, in Gaia DR4 (which will contain 5 years of data) we expect 640 microlensing events and in a potential Gaia DR5 (containing 10.5 years of data) we expect 1300 microlensing events.

We can estimate the uncertainty of raw positions that will appear in DR4/5 using the recent Gaia focused product release which contains the observed raw position for quasars \citep{Krone-Martins2023}. By calculating the centroid of each source we reach a standard deviation $\sim 50$~mas in each individual measurement. A typical Gaia source is observed $\sim 20$ times per year for a total of 210 measurements over the full 10.5 years of observation. Assuming the uncertainty scales with the square root of the number of observations, the uncertainty of a typical object analysed in isolation will be $\sim 3$~mas --- much too large to detect astrometric microlensing events which are typically 5~mas at most. This result highlights the necessity of including a microlensing model into the astrometric solution of the Gaia data analysis pipeline. With a microlensing model included, astrometric microlensing events should be detectable below the 3~mas level, as demonstrated by the tight astrometric solutions the existing Gaia pipeline provides for sources.

A similar analysis applied to the Roman Galactic Bulge Time-Domain Survey which is currently expected to achieve 1~mas uncertainty on individual astrometric measurements with $\sim 7000$ measurements made over 6 observing seasons results in astrometric microlensing events being detectable as low as 20~$\mu$as. If achieved, this would result in colossal numbers of microlensing events being detected.

\subsection{Predictions for GaiaNIR}
\begin{figure}
	\includegraphics[width=\columnwidth]{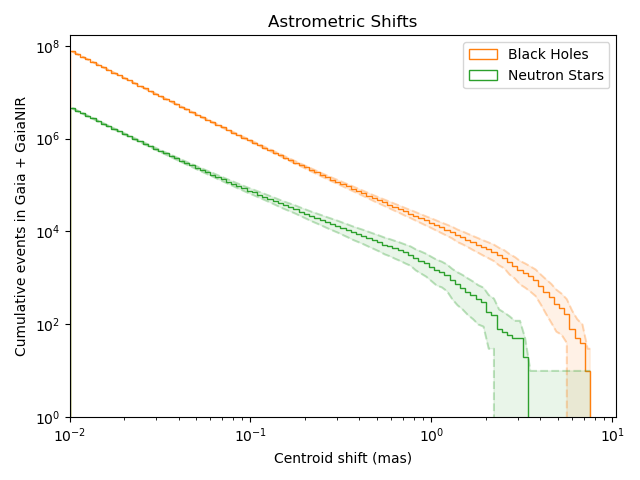}
	\includegraphics[width=\columnwidth]{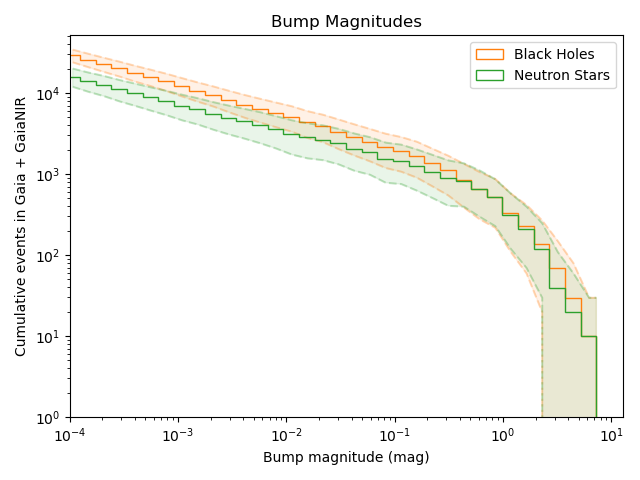}
	\includegraphics[width=\columnwidth]{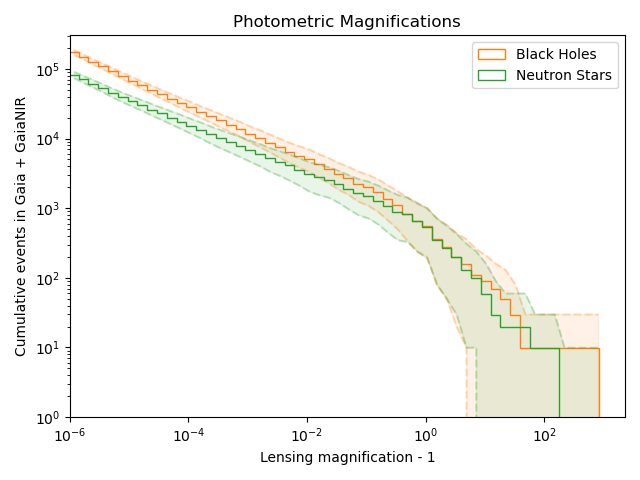}
    \caption{Expected distribution for GaiaNIR of astrometric shifts, bump magnitudes and lensing magnifications caused by microlensing events where the lenses are BHs and NSs. The shaded regions indicate $3\sigma$ uncertainty. Note that in the magnification plot the $x$ axis is (magnification - 1), permitting a log scale. The relationships will continue below the 10$^{1}$ threshold, but are below the sensitivity of our simulation.}
    \label{fig:GaiaNIR}
\end{figure}

In addition to comparing our results to existing Gaia data, we can make predictions for microlensing events observed by the combination of Gaia and GaiaNIR data. As described in Section~\ref{sec:timespans}, \textit{Simulation 3} contains 101 ``observation runs'' based on an observing window of 12 years (to mimic the Gaia observation window), then a break of 15 years followed by another observing window of 5 years intended to emulate GaiaNIR observations (assuming GaiaNIR begins observing in 2041). By performing 101 such simulations we undersample the true distribution by a factor of $\sim$10, since the distribution from \citet{Sweeney2022} undersamples by a factor of 1\,000.

The predicted distributions of centroid shifts, bump magnitudes and photometric magnifications are shown in Figure~\ref{fig:GaiaNIR}. These results show that the combination of Gaia and GaiaNIR data is likely to contain $4200_{-400}^{+400}$ BH and $190_{-80}^{+90}$ NS microlensing events with an astrometric shift larger than 2~mas ($14700_{-900}^{+600}$ BH and $1600_{-200}^{+300}$ NS events larger than 1~mas). Similarly, there are likely to be $330_{-120}^{+100}$ BH and $310_{-100}^{+110}$ NS events with bump magnitudes $> 1$~mag and $90_{-60}^{+60}$ and $60_{-40}^{+50}$ BH and NS events larger with bump magnitudes $< -2.5$~mag. If we use half the number of events with bump magnitude $\geq 3$~mag (lensing magnification $\geq 15$), G magnitude $< 19$ and $t_E > 20$~days as a proxy for \citet{Wyrzykowski2023}'s selection criteria then the same analysis is likely to detect $<10$ events from BHs and $5_{-5}^{+10}$ from NSs in the combination of Gaia and GaiaNIR data. Note that these numbers are optimistic as the approximation of the \citet{Wyrzykowski2023} selection function is for their optimal detection efficiency, not the entire sky.

By contrast, our yearly results (Section~\ref{sec:yearly}) would estimate only a fraction of the astrometric event counts. This apparent surplus of microlensing events in our combination of Gaia and GaiaNIR data is due to our simulation ``detecting'' many events which do not peak during the observation windows. 
In the analysis by \citet{Wyrzykowski2023} they were able to detect events which peaked outside their observation period so we expected many of these events with longer Einstein times will be recoverable.
The observation windows also lead to more 1~mas deflection BH events being detected in comparison to $1$~mas deflection NSs events, $9.2\times$ more as opposed to $4.2\times$ more for events peaking in a typical year. This increase in the fraction of BHs being detected is because of their generally longer Einstein times, so there are more ``heads'' and ``tails'' of BH events occurring at the start/end of any given observation period. 

\begin{figure}
	\includegraphics[width=\columnwidth]{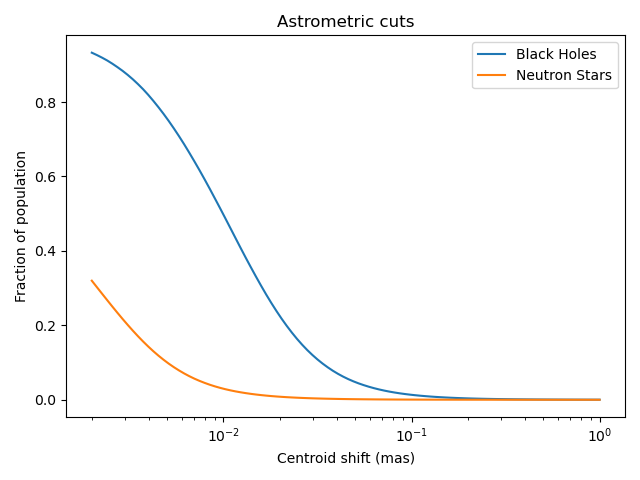}
	\includegraphics[width=\columnwidth]{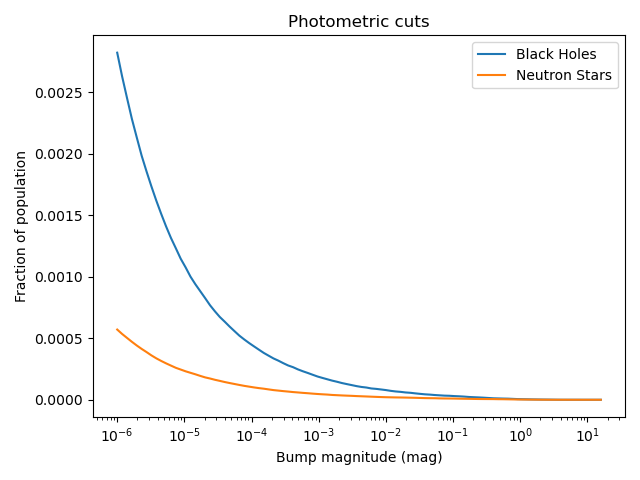}
    \caption{Figure showing the fraction of the BH or NS population which experiences at least one microlensing event larger than an astrometric or photometric threshold in the combination of Gaia and GaiaNIR data.}
    \label{fig:lensing-fraction}
\end{figure}

The fraction of the BH or NS population which experiences at least one microlensing event is a function of both the observing period and the detection criteria for a microlensing event. Figure~\ref{fig:lensing-fraction} shows this lensing fraction for BHs and NSs as a function of centroid shift and bump magnitude for the combination of Gaia and GaiaNIR data. From this figure we can see that in order to probe a reasonable fraction of the isolated BH or NS population we must be able to identify lensing events which cause small astrometric shifts.

\section{Conclusions}

This paper has explored the rates at which microlensing events occur across the entire sky. This was modelled on both a yearly basis and for the ongoing Gaia mission combined with a future GaiaNIR mission. Microlensing event rates were calculated by performing an N-body simulations using the BH and NS population published by \citet{Sweeney2022} and a stellar population generated by {\tt GALAXIA} as the lensing populations. The population of BGSs was drawn from the Gaia catalogue so the selection function of stars monitored for microlensing events would be highly realistic.

The key methodological advancements in this work are (1) the population of NSs and BHs modelled with realistic natal kicks and subsequently evolved through the Galactic potential and (2) the use of Gaia stars as BGSs to ensure a realistic selection function. The results of these simulations were compared to microlensing events detected in Gaia data \citep{Wyrzykowski2023} and were found to be consistent with these observations. Our main findings are as follows:

\begin{itemize}
    \item On a yearly basis we can expect $88_{-6}^{+6}$ BH, $6.8_{-1.6}^{+1.7}$ NS and $20^{+30}_{-20}$ stellar microlensing events which cause a blended astrometric shift larger than 2~mas. For events causing an astrometric shift larger than 1~mas, we expect $231_{-9}^{+10}$ BH events, $55_{-5}^{+5}$ NS events and $690_{-160}^{+170}$ stellar events. Similarly, we can expect $21_{-3}^{+3}$ BH, $18_{-3}^{+3}$ NS and $7500_{-500}^{+500}$ stellar microlensing events with bump magnitudes larger than 1~mag and $5.5_{-1.4}^{+1.5}$, $4.3_{-1.2}^{+1.3}$ and $1600_{-200}^{+300}$ events with bump magnitudes larger than 2.5.
    \item Our results update the BH and NS event statistics from \citet{PopSyCLE} who found that events with $t_E > 120$~days have a 50\% probability of being a BH. We find that when the BHs are evolved post-natal-kick to the present day the fraction of BH events at $t_E \approx 120$~days is much lower, due to both an increase in the Einstein times of BH events as well as a decrease in the number of BH events. Here we find that 28\% of $t_E > 1\,000$~day events are caused by BHs if events are selected by a bump magnitude cut. This fraction is larger if events are selected by an astrometric cut.
    \item We also consider the scenario where BHs have a mass of 20~M$_\odot$, but receiving the same natal kick. This leads to BHs causing $\sim $50\% more photometric microlensing events, still mostly at very long Einstein times.
    \item In agreement with \citet{PopSyCLE} and \citet{Gould2023}, we find that measurements of both $\pi_E$ and $t_E$ are required to identify BH events.
    \item We provide distributions of lens masses and lens distances broken down by species that may be valuable as informative priors for observational data analysis.
    \item The microlensing events found in Gaia data \citep{Wyrzykowski2023} are consistent with our data. Considering mid-length ($20 < t_E < 250$) bulge events ($|l| < 10$) for bright ($G < 19$~mag) Gaia stars, emulating the filters in \citet{Wyrzykowski2023}'s analysis, we predict 100 yearly events with bump magnitudes $> 3$~mag compared to \citet{Wyrzykowski2023}'s $\sim 50$ yearly detected events.
    We find that the distribution of Einstein times and their longitude also show very good agreement. Our data is symmetrical around a 0 Galactic latitude, whereas \citet{Wyrzykowski2023}'s analysis is not. We conclude that this is due to the selection bias in the OGLE-IV survey upon which their analysis depends on for candidate selection.
    \item We predict 640 microlensing events will be detected in the photometry of Gaia DR4 and 1300 microlensing events in Gaia DR5. We use a recent Gaia focused product release \citep{Krone-Martins2023} to estimate the number of astrometric events which will be recoverable from the raw Gaia positions to be released in DR4/5 and conclude that the uncertainties will be too large to detect any events unless a microlensing model is included into the Gaia data analysis pipeline. We briefly consider astrometric events which will be detectable in the Roman Galactic Bulge Time Domain Survey and conclude that if they achieve the expected 1~mas uncertainty on individual measurements then microlensing events as small as 20~$\mu$as may be detectable.
    \item We predict $4200_{-400}^{+400}$ BH and $190_{-80}^{+90}$ NS microlensing events will be present in the combination of Gaia and GaiaNIR data with centroid shifts larger than 2~mas ($14700_{-900}^{+600}$ BH and $1600_{-200}^{+300}$ NS events larger than 1~mas). Similarly, we predict $330_{-120}^{+100}$ BH and $310_{-100}^{+110}$ NS events with bump magnitudes $> 1$~mag ($90_{-60}^{+60}$ BH and $60_{-40}^{+50}$ NS events $> 2.5$~mag). Using half the number of events with bump magnitudes $> 3$~mag, G magnitude $< 19$ and $t_E > 20$~days as a proxy for \citet{Wyrzykowski2023}'s selection function, we predict that $<10$ BH and $5_{-5}^{+10}$ NS events will be detectable in the combination of Gaia and GaiaNIR data if the same analysis is performed.
\end{itemize}

The recent detection of an isolated BH \citep{Lam2022, Sahu2022} has demonstrated that it is possible to peer into the hidden world of dark compact remnants. Observing these electromagnetically invisible objects provides a probe for massive binary evolution and supernovae through Galactic history. We characterise these signatures as precisely as modern theory allows and make predictions against which the observational community can test future measurements. In particular, the BH and NS events are strongly dependant on the natal kicks received during supernovae and the masses of the compact remnants. \citet{Sweeney2022} established that 40\% of NSs are ejected from the Galaxy due to these natal kicks, which significantly reduces the number of microlensing events expected from NSs. Sufficient statistics from future studies would allow constraints to be placed on these natal kicks and masses, both of which are uncertain --- especially in the case of BHs.

\section*{Acknowledgements}

We would like to thank Subo Dong, Tim Bedding, James Crowley, Jessica Lu, Natasha Abhrams, Casey Lam and Lukasz Wyrzykowski for their helpful conversations. This collaboration work was in part made possible by ESO's early-career scientific visitor programme. AKM acknowledges partial support from the Portuguese Funda\c c\~ao para a Ci\^encia e a Tecnologia project EXPL/FIS-AST/1368/2021. This work has made use of data from the European Space Agency (ESA) mission {\it Gaia} (\url{https://www.cosmos.esa.int/gaia}), processed by the {\it Gaia} Data Processing and Analysis Consortium (DPAC, \url{https://www.cosmos.esa.int/web/gaia/dpac/consortium}). Funding for the DPAC has been provided by national institutions, in particular the institutions participating in the {\it Gaia} Multilateral Agreement. In addition to those mentioned in the main text, we would also like to acknowledge other python packages used as part of this work: NumPy \citep{NumPy}, Matplotlib \citep{Matplotlib}, seaborn \citep{seaborn}, pandas \citep{pandas2010, pandas}, Astropy \citep{astropy2013, astropy2018, astropy2022}, astroquery \citep{astroquery} and ray \citep{ray}.

\section*{Data Availability}

The data and code underlying this article are available on Zenodo \citep{zenodo-code}, as well as at the following GitHub url: \url{https://github.com/David-Sweeney/Incidence}.



\bibliographystyle{mnras}
\bibliography{example} 



\appendix

\section{Gaia null values}
\label{app:null-Gaia}

The Gaia catalogue has null values scattered throughout, as you would expect from a large survey. If our simulation queries the Gaia catalogue and receives a null value we replace it with a value of 0 if the null value is a parallax or radial velocity. If the null value is a proper motion in right ascension then the value is set to  $-1.885932688110818$; a proper motion in declination is set to $-3.9745764127030273$. These proper motion values were chosen as they were the median values of some early trial runs. We do not replace null values in the G magnitude column. Where they exist the null values propagate through to the blended centroid shift and bump magnitude columns (making these values null).

\section{BH event fraction}
\label{app:fraction}

\begin{figure}
	\includegraphics[width=\columnwidth]{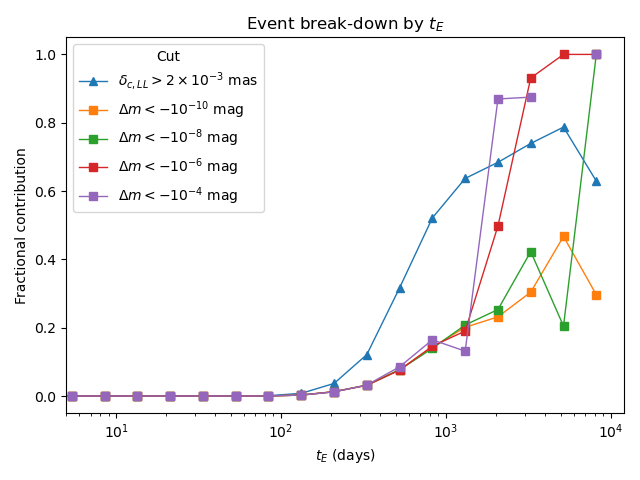}
    \caption{Fraction of events caused by BHs as a function of Einstein time for an astrometric cut and various bump magnitude cuts. These lines are the counterpart of the orange line in Figure~\ref{fig:einstein-times} depending on the selected events.}
    \label{fig:photometric-cuts}
\end{figure}

\begin{figure}
	\includegraphics[width=\columnwidth]{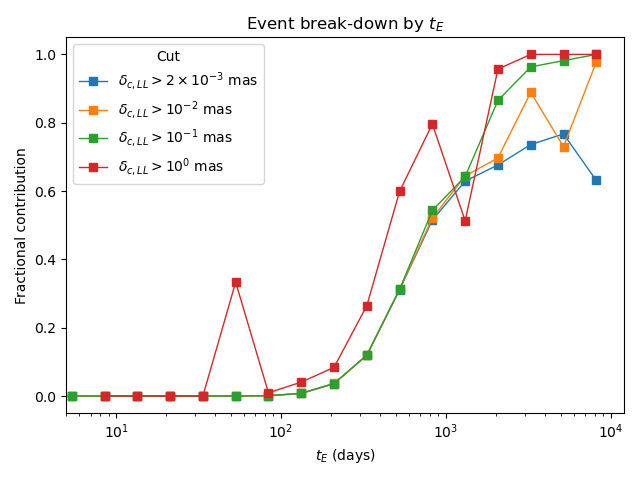}
    \caption{Fraction of events caused by BHs as a function of Einstein time for various astrometric cuts. These lines are the counterpart of the orange line in Figure~\ref{fig:einstein-times} depending on the selected events.}
    \label{fig:astrometric-cuts}
\end{figure}

Plotted in Figure~\ref{fig:photometric-cuts} and Figure~\ref{fig:astrometric-cuts} are the fraction of events which are caused by BHs as a function of time for various selection criteria applied to the population. We can see that the threshold of the cut does not affect the shape of the contribution curve, if the noise at high Einstein times is ignored (where the number of stellar events is extremely low due to the undersampling of the stellar population). Instead, the shape of the curve is mainly determined by the type of cut applied: photometric or astrometric. For microlensing events collected from photometric surveys the flatter curve of BH contribution is expected.

\section{Einstein time variation in and outside the bulge}
\label{app:bulge}

\begin{figure}
	\includegraphics[width=\columnwidth]{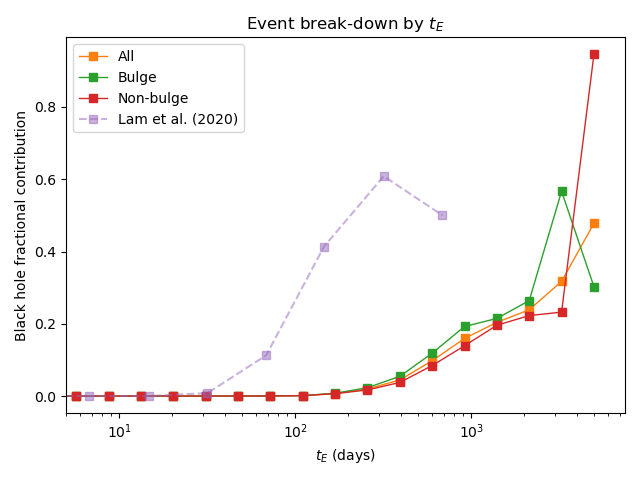}
    \caption{Fraction of yearly microlensing events caused by BHs as a function of Einstein time, split by population and compared to the results from \citet{PopSyCLE}. The largest bin/point in this figure was set to be $6\,000$~days so the larger stellar undersampling doesn't mislead the viewer. BH, NS and some stellar events continue beyond this point. This figure includes events down to a bump magnitude of $10^{-10}$~mag, many of which will not be observable.}
    \label{fig:bulge-fraction}
\end{figure}

Defining the bulge as $|l| < 30$~deg and $|b| < 10$~deg, we can plot the fraction of event which have BH lenses as a function of Einstein time for the bulge, non-bulge and total population, as shown in Figure~\ref{fig:bulge-fraction}. This figure shows that there is only a slight difference in the BH fraction inside/outside the bulge. For completeness we also plot the results from \citep{PopSyCLE}, ignoring their events due to white dwarfs (which they calculate while we don't).

\begin{figure}
	\includegraphics[width=\columnwidth]{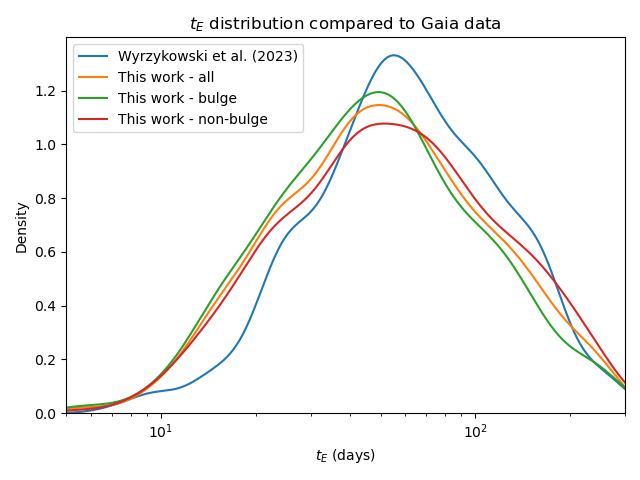}
    \caption{Comparison between the distribution of Einstein times found in Gaia data \citep{Wyrzykowski2023} and those reached in this work, split by whether they were/were not in the bulge. 
    Microlensing events due to stars dominate the distributions from this work.
    The distribution from this work only includes events with a bump magnitude $>0.4$~mag in order to emulate a photometric selection criteria. The Gaia data plotted is the ``Level 1'' solutions derived in \citet{Wyrzykowski2023}. Some outliers from the Gaia data are beyond the bounds of this plot. The curves were generated using a Gaussian kernel density estimator, so each curve will integrate to 1.}
    \label{fig:bulge-times}
\end{figure}

We also plot the distribution of Einstein times for events inside and outside the bulge, as shown in Figure~\ref{fig:bulge-times}. We overplot this on Figure~\ref{fig:Gaia-data} which shows the distribution for our entire population and the distribution from \citet{Wyrzykowski2023}.

\section{Distribution of microlens parallax}
\label{app:microlens-parallax}

\begin{figure}
	\includegraphics[width=\columnwidth]{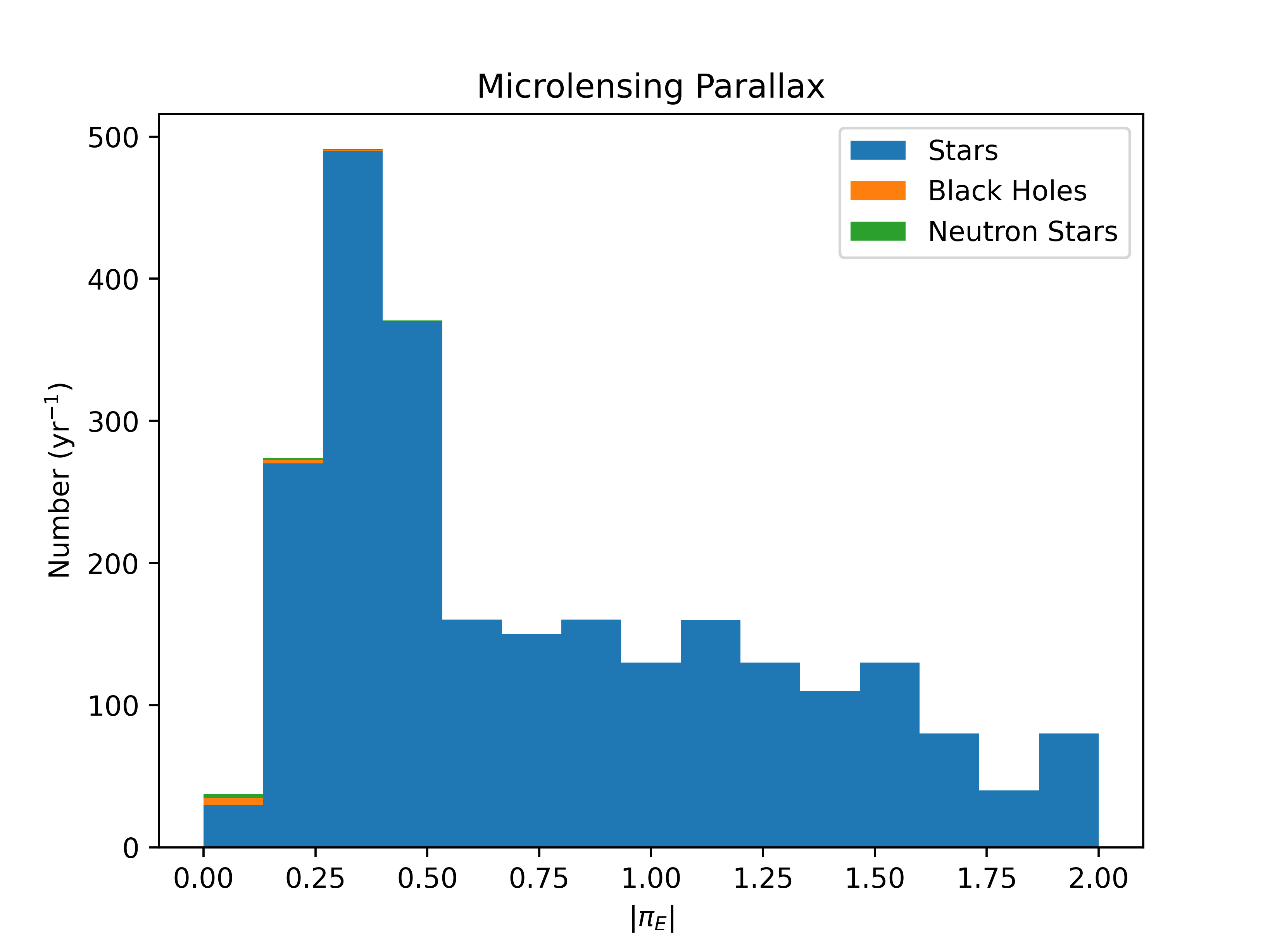}
    \caption{Expected yearly distribution of microlens parallax, shown for events down to a bump magnitude of $2$~mag. The largest bin in this figure was set to be 2~mas, some events do occur beyond this point.}
    \label{fig:microlens-parallax-hist}
\end{figure}

Plotted in Figure~\ref{fig:microlens-parallax-hist} is the yearly distribution of microlens parallax. We can see from the figure that stellar events dominate the distribution at all values of $|\pi_E|$, but that BH and NS events are most present at small values of $|\pi_E|$. Note the uncertainty in the Gaia measurement for parallax often results in a negative parallax value. For this reason we plot the absolute value of $\pi_E$.

\section{Event rates for 20~M$_\odot$ black holes}
\label{app:20M-BH}
If the population of BHs have a mass of 20~M$_\odot$ then they cause $670_{-19}^{+19}$ and $268_{-11}^{+11}$ events with centroid shifts larger than $1$~mas and $2$~mas, respectively. They cause $32_{-4}^{+4}$ and $8.1_{-1.7}^{+1.8}$ events with bump magnitudes $> 1$ and $> 2.5$, respectively. For comparison, with a mass of 7.8~M$_\odot$ the respective numbers are $231_{-9}^{+10}$, $88_{-6}^{+6}$, $21_{-3}^{+3}$ and $5.5_{-1.4}^{+1.5}$ (in the same order). Thus, the $2.6\times$ increase in BH mass ($1.6\times$ increase in $\theta_E$) leads to a $\sim 3\times$ increase in astrometric signals and a $\sim 1.5\times$ increase in photometric signals.

\section{Distribution of lens distances and masses}
\label{app:lens-info}

\begin{figure}
	\includegraphics[width=\columnwidth]{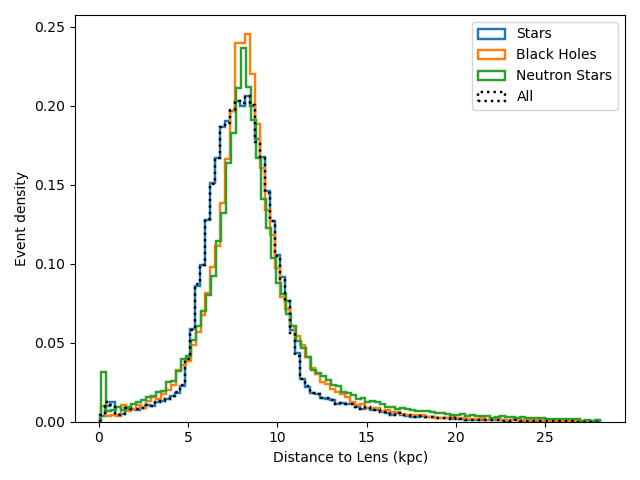}
    \caption{Distribution of lens distances for microlensing events caused by different populations. The distribution does not change significantly for different astrometric or photometric cuts. This figure includes events down to a major image shift $>1$~$\mu$as, many of which will not be observable.}
    \label{fig:lens-distance}
\end{figure}

\begin{figure}
	\includegraphics[width=\columnwidth]{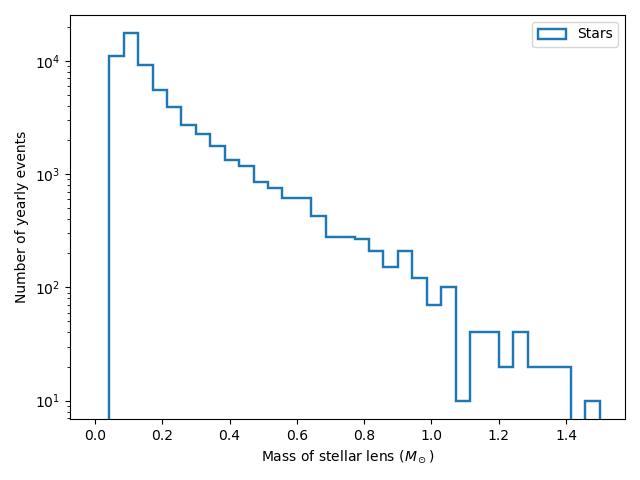}
    \caption{Distribution of lens masses for stellar lenses. There is a mild dependence on the photometric cut applied, so this figure includes events with bump magnitude $> 0.01$~mag. The distribution is not significantly different for events with bump magnitude $> 0.1$~mag. The mass distribution for BHs and NSs is not plotted as they were all taken to have a mass of 7.8~M$_\odot$ and 1.35~M$_\odot$, respectively.}
    \label{fig:lens-mass}
\end{figure}


\bsp	
\label{lastpage}
\end{document}